\documentclass[useAMS]{mn2e}
\usepackage{epsfig}
\usepackage{graphicx}

\newcommand{\degree}[1]{\ensuremath{#1^{\circ}}}

\begin{document}

\title[Radiation-driven warped discs]
  {Three dimensional SPH simulations of radiation-driven warped accretion discs}
\author[S. B. Foulkes et al.]
  {Stephen B. Foulkes$^{1}$, Carole A. Haswell$^{1}$, James R. Murray$^{2}$ \\
$^1$Department of Physics \& Astronomy, The Open University, Walton Hall, Milton Keynes, MK7 6AA, UK. \\
$^2$Department of Astrophysics \& Supercomputing, Swinburne University of Technology, Hawthorn, VIC 3122, Australia.\\
Email SBF: sbfoulkes@qinetiq.com, CAH: C.A.Haswell@open.ac.uk, JRM: jmurray@astro.swin.edu.au }

\date{Accepted. Received}
\pagerange{\pageref{firstpage}--\pageref{lastpage}}
\pubyear{2005}

\maketitle \label{firstpage} 
\begin{abstract}
We present three dimensional smoothed particle hydrodynamics (SPH) calculations of warped accretion discs in X-ray binary systems. Geometrically thin, optically thick accretion discs are illuminated by a central radiation source. This illumination exerts a non-axisymmetric radiation pressure on the surface of the disc resulting in a torque that acts on the disc to induce a twist or warp. Initially planar discs are unstable to warping driven by the radiation torque and in general the warps also precess in a retrograde direction relative to the orbital flow. We simulate a number of X-ray binary systems which have different mass ratios using a number of different luminosities for each. Radiation-driven warping occurs for all systems simulated. For mass ratios q $\sim 0.1$ a moderate warp occurs in the inner disc while the outer disc remains in the orbital plane (c.f. X$\thinspace 1916-053$). For less extreme mass ratios the entire disc tilts out of the orbital plane (c.f. Her X-1). For discs that are tilted out of the orbital plane in which the outer edge material of the disc is precessing in a prograde direction we obtain both positive and negative superhumps simultaneously in the dissipation light curve (c.f. V603 Aql).   
\end{abstract}

\begin{keywords}
  accretion: accretion discs -
  X-rays: binaries -
  binaries: close X-ray binaries- 
  methods: numerical.
\end{keywords}

\section{Introduction}
There are a growing number of identified X-ray binaries that have periodic variabilities in their light curves that are significantly longer than their binary orbital periods. These `super-orbital' periods have been found in Her X-1, SS 433 and LMC X-4 for example \cite{ClarksonEt:2003}. A number of different explanations have been proposed for these super-orbital periodicities, including the precession of a tilted or warped accretion disc (Katz 1973; Petterson 1975; Wijers \& Pringle 1999 and references therein). X-rays from close to the accreting object are reprocessed and re-emitted at the surface of the accretion disc. As the disc presents a changing aspect toward the observer, so light variations are seen. Variations in the amount of the donor star eclipsed by the warped disc and the different X-ray illumination of the donor (Gerend \& Boynton 1976) also generate modulations in the light curves. A tilted disc will also obscure the central X-ray source from the observer and generate `dips' in the X-ray light curves as seen for example in X1916-053 \cite{HomerEt:2001}.

Tananbaum et al. (1972) interpreted the 35-day period in the X-ray flux from Her X-1 as resulting from an accretion disc that is precessing and tilted with respect to the orbital plane. Katz (1973) suggested that the precession was forced by a torque from the donor star, although the mechanism for the misalignment of the disc out of the orbital plane was unclear. Other mechanisms have been suggested for driving the warps such as wind torques \cite{SchandlMeyer:1994}.

The discovery of the precessing relativistic jets found in SS 433 (see Margon 1984 for a comprehensive review) indicated that an accretion disc was present which was precessing with a period of 164 days. Moreover, the morphology of the radio jets and optical photometry of the system, in combination with the jet precession, all indicate that the entire disc precesses at the same rate. Global precession of the disc is also evident for Her X-1 (Tr\"umper et al. 1986; Petterson et al. 1991).

Radiation pressure warping of an initially coplanar accretion disc was first detailed by Petterson (1977). Iping \& Petterson (1990) performed detailed numerical simulations of warped discs, suggesting that radiation pressure determines the shape of the disc and its precession rate. If radiation from the central source is absorbed at the surface of the accretion disc and then re-emitted normal to this surface, a torque will be induced on the disc (Petterson 1977 ; Pringle 1996). For values of the central radiation luminosity sufficient to overcome the viscous forces within the disc, a warp will develop and the disc will tilt out of the orbital plane (Pringle 1996). 

Pringle (1996, 1997) investigated analytically and numerically the effects of radiation-warping on an initially flat disc using a linear analysis for isothermal $\alpha$-discs. The disc was modelled using a set of concentric rings centred on the primary object. Neighbouring rings exchanged angular momentum by means of viscous torques. Mass and angular momentum were conserved, but there was no detailed consideration of the internal fluid dynamics of the disc.  Pringle concluded that an accretion disc is unstable to tilting and warping due to irradiation reaction forces when the luminosity of the central source exceeds a critical value. Wijers \& Pringle (1999) extended Pringle's work and addressed the non-linear evolution of the instability and found very different and complex warp development. 

Larwood et al. (1996) investigated tidally induced warping and precession of accretion discs in binary systems. They used a smoothed particle hydrodynamics (SPH) code to investigate how a thin accretion disc that is warped and inclined to the orbital plane can survive in close binary systems. They were able to demonstrate disc warping and precession in the tidal field of the companion. The disc was only moderately warped and precessed approximately like a rigid body. A similar study of warped circumbinary discs was performed by Larwood \& Papaloizou (1997). Nelson \& Papaloizou (1999) studied the dynamics of warped discs orbiting a rotating black hole where the disc and black hole angular momentum vectors where misaligned. In this case the inner regions of the disc became warped.   

In these various investigations a number of different types of disc warp have been found, including stable, unstable and time-independent modes. In most cases the warp precesses relative to the inertial frame. The precession rate is generally constant and the warp precesses as a rigid body by twisting the disc such that the radiation pressure causes each radial annulus to precess at the same rate. The rate of growth of the warp is dependent on the outer disc boundary conditions. The warp generally starts in the outer parts of the disc and propagates inward.

Murray et al. (2002) modelled an initially planar disc subject to an inclined magnetic dipole field centred on the donor star. They used a SPH code to model the disc response to a magnetic field that was fixed in the binary frame. They found a warp developed that precessed uniformly in a retrograde direction relative to the direction of the disc flow. The amplitude and structure of the warp was dependent on the relative phase with respect to the magnetic field. The induced warp was at maximum amplitude when it was reinforced by the magnetic force and a minimum when the warp and magnetic force were anti-phased, giving an almost flat disc.

In this paper we present a numerical study of radiation-driven warping for a number of X-ray binary systems. We investigate the problem using a non-linear SPH code originally developed by Murray (1996, 1998). In particular we aim to establish to what extent a thin, initially coplanar disc will warp and precess subject to irradiation from a central source.  In section \ref{sec:Basic Equations} and \ref{sec:Numerical method} we present our formulation of the problem using SPH. Section \ref{sec:Simulations} details the binary systems modelled, and in Section \ref{NUMERICAL RESULTS} we present our results. Section \ref{sec:Conclusions} is devoted to conclusions.

\section{Basic Equations}
\label{sec:Basic Equations}
We take the basic hydrodynamics equations governing the viscous gaseous material of the accretion disc. It is assumed that the disc self-gravity is negligible in comparison to the binary component's tidal field. The equations of continuity and motion may be written as

\begin{equation}\label{eq:1}
\frac {D \rho}{D t} + \rho \mathbf{\nabla} \cdot \mathbf{v} = 0
\end{equation}

\begin{equation}\label{eq:2}
\frac {D \mathbf{v}}{D t} =  -\frac {1} {\rho} \mathbf{\nabla} P - \mathbf{\nabla} \Phi + \mathbf{f_{v}} + \mathbf{f_{irr}}
\end{equation}

\noindent where in the above

\begin{equation}\label{eq:2a}
\frac {D }{D t} \equiv  \frac {\partial}{\partial t} + \mathbf{v} \cdot \mathbf{\nabla}
\end{equation}  

\noindent is the material derivative operator (Batchelor 1970), $\rho$, $\mathbf{v}$ and $P$ are the fluid density, velocity and pressure values respectively. The gravitational potential is $\Phi$, the viscous force is $\mathbf{f_{v}}$ and the force due to the radiation on the fluid is $\mathbf{f_{irr}}$. 
The equation for the rate of change of thermal energy per unit mass can be expressed as
\begin{equation}\label{eq:2c}
\frac {d u}{d t} =  - \left ( \frac {P}{\rho} \right ) \mathbf{\nabla} \cdot \mathbf{v}.
\end{equation}  

\section{Numerical method}
\label{sec:Numerical method}
We solve the basic equations in section \ref{sec:Basic Equations} using a SPH (Lucy 1977, Gingold \& Monaghan 1977) computer code first developed by Murray (1996, 1998).

\subsection{SPH formulation}
SPH uses smoothed quantities defined by the equation
\begin{equation}\label{eq:3}
\langle f\left (\mathbf{r}\right) \rangle = \int f(\mathbf{r'})W\left( \vert \mathbf{r-r'}\vert, h\right )d\mathbf{r'},
\end{equation}  

\noindent here $f(\mathbf{r})$ is an arbitrary function, $W$ is a smoothing kernel, $h$ is the smoothing length and the integration is over the entire space.  A Monte-Carlo representation of $(\ref{eq:3})$ is used for each particle position

\begin{equation}\label{eq:4}
\langle f\left (\mathbf{r_{i}}\right) \rangle = \sum_{j=1}^N \frac{m_{j}}{\rho_{j}}f(\mathbf{r_{j}})W\left( \vert \mathbf{r_i-r_j}\vert, h\right ),
\end{equation} 

\noindent where $m_j$ is the mass of particle $j$, $\rho_j$ is the density of the particle $j$ and the sum is over all particles N. We use a variable smoothing length such that each particle, $i$, has an associated smoothing length $h_i$. The kernel is symmetrized  with respect to particle pairs in order to conserve momentum.

\subsubsection{Equations of motion}
The equation of motion for each particle $i$ is given by
\textsl{}
\begin{equation}\label{eq:5}
m_i\frac{d\mathbf{v}_i}{dt}=\mathbf{F}_{P,i}+\mathbf{F}_{G,i}+\mathbf{F}_{visc,i}+\mathbf{F}_{irr,i},
\end{equation} 

\noindent which gives the force on particle $i$ as a summation of the pressure, gravity, viscous and irradiation forces, $\mathbf{F}_{P,i}$, $\mathbf{F}_{G,i}$, $\mathbf{F}_{visc,i}$ and $\mathbf{F}_{irr,i}$ respectively.

\subsubsection{The pressure term}
For the pressure gradient we make use of the fact that

\begin{equation}\label{eq:6}
\frac{1}{\rho}\mathbf{\nabla} P = \mathbf{\nabla} \left ( \frac{P}{\rho} \right ) + \frac{P}{\rho^2} \mathbf{\nabla} \rho,
\end{equation} 

\noindent and the pressure is given by

\begin{equation}\label{eq:7}
\left ( \frac{1}{\rho}\mathbf{\nabla}P \right )_i = \sum_{j} m_j \left (\frac{P_i}{\rho_i^2} + \frac{P_j}{\rho_j^2} \right) \mathbf{\nabla}W \left(\vert \mathbf{r_i} - \mathbf{r_j} \vert, h \right ) .
\end{equation} 

\noindent Similarly the SPH energy equation is 
\begin{equation}\label{eq:7a}
\frac{du_i}{dt} = \frac{1}{2} \sum_{j} m_j \left (\frac{P_i}{\rho_i^2} + \frac{P_j}{\rho_j^2} \right)\mathbf{v_{ij}} \cdot \mathbf{\nabla} W \left(\vert \mathbf{r_i} - \mathbf{r_j} \vert, h \right ) .
\end{equation} 

\subsubsection{The artificial viscosity term}
To stabilise the shock flow regions we used an artificial viscosity term as detailed by Murray (1996) and Truss et al. (2000). An artificial viscosity term was added to the pressure term such that

\begin{equation}\label{eq:8}
\frac{P_i}{\rho_i^2} + \frac{P_j}{\rho_j^2} \rightarrow \frac{P_i}{\rho_i^2} + \frac{P_j}{\rho_j^2} + \Pi_{ij}
\end{equation} 

\noindent where the artificial viscous pressure is given by the Gingold \& Monaghan (1983) equation

\begin{equation}\label{eq:9}
\Pi_{ij} = \frac{1}{\bar\rho_{ij}}\left(-\alpha\mu_{ij}\bar c_{ij}  + \beta\mu_{ij}^2 \right).
\end{equation} 

\noindent Here we are using the notation $\bar A_{ij} = \frac{1}{2}\left(A_i+A_j\right)$ and $\bar c_{ij}$ is the average sound speed at particle positions $i$, $j$ and


\begin{equation}\label{eq:10}
\mu_{ij} = \bar h_{ij}\frac{\mathbf{v_{ij}}\cdot \mathbf{r}_{ij}}{\mathbf{r}_{ij}^2+\eta^2}. 
\end{equation} 

\noindent In this equation $\mathbf{v}_{ij} = \mathbf{v}_i-\mathbf{v}_j$ and $\eta^2=0.01\bar h_{ij}^2$ which prevents the denominator going to zero. We used $\alpha = 0.5$ and $\beta=0$ for the simulations reported here.  

\subsubsection{The radiation force}
 We follow the geometrical definitions of Pringle (1996) and define a unit tilt vector (see Fig. 1 of Ogilvie 1999) $\mathbf{l}(R,t)$ such that

\begin{equation}\label{eq:11}
\mathbf{l}=(\cos\gamma \sin\beta, \sin\gamma \sin \beta, \cos \beta)
\end{equation} 

\noindent where $\beta(R,t)$ and $\gamma(R,t)$ are the Euler angles of the tilt vector with respect to a fixed Cartesian coordinate system $(OXYZ)$ centered on the compact object.
The position vector of a point on the disc with distance $R$ from the origin and azimuth $\phi$ is given by $R\mathbf{e_{R}}$ where $\mathbf{e_{R}}$ is the radial unit vector and is given by

\begin{equation}\label{eq:12}
\mathbf{e_{R}}=\left[ \begin{array}{c}
\cos\phi \sin\gamma + \sin\phi \cos\gamma \cos \beta \\
-\cos\phi \cos\gamma + \sin\phi \sin\gamma \cos\beta \\
-\sin\phi \cos\beta 
\end{array} \right] .
\end{equation} 

\noindent The radiation source is assumed to be centered on the compact object position and radiates isotropically. The radiation flux at a distance $R$ from the compact object is given by

\begin{equation}\label{eq:13}
\mathbf{f}=\frac{L_{*}}{4\pi R^2} \mathbf{e_{R}},
\end{equation} 
and the power absorbed at a surface element $d\mathbf{S}$ is

\begin{equation}\label{eq:14}
dP= \left (1-\eta \right ) \big | \mathbf{f} \cdot d\mathbf{S} \big | ,
\end{equation} 

\noindent where $L_{*}$ is the total luminosity of the radiation source, $\eta$ is the disc albedo, that is the amount of radiation scattered by the surface of the disc, in all the simulations reported in this paper $\eta$ was assumed to be zero, and $d\mathbf{S}$ is an element of the disc surface area given by

\begin{equation}\label{eq:15}
d\mathbf{S}=\left(\frac{\partial \mathbf{R}}{\partial R}dR \right) \times \left(\frac{\partial \mathbf{R}}{\partial \phi}d\phi \right).
\end{equation} 

\noindent We assume that the radiation is absorbed by this element and is uniformly re-radiated immediately and on the same disc side. The force due to absorption of the radiation is

\begin{equation}\label{eq:16}
d\mathbf{A}= \frac{dP}{c} \mathbf{e_R},
\end{equation} 

\noindent where $c$ is the speed of light. 

Since the incident radiation is parallel to the radius vector a circular disc annulus receives no torque from the absorbed radiation. There is, however,  a net torque on a circular annulus due to the summed radiation pressure reaction, as calculated by Pringle (1996).  Since the radiation is re-radiated uniformly from the disc surface, the element will receive a radiation pressure reaction from this radiation of the form

\begin{equation}\label{eq:17}
d\mathbf{F}= -\frac{2}{3}\frac{dP}{c}\mathbf{n},
\end{equation} 

\noindent where $\mathbf{n}$ is the unit normal pointing away from the disc surface that received the initial radiation.

\section{Simulations} 
\label{sec:Simulations}  
\subsection{Binary Parameters}
The accretion disc was modelled using a three-dimensional SPH computer code that has been described in detail in Murray (1996, 1998) and Truss et al. (2000), modifications have been made to convert the code from FORTRAN to C++.  The SPH results presented here were generated using The Swinburne Centre for Astrophysics and Supercomputing facility and a local private Linux system. 

In each simulation the total system mass, $M_{t}$, and the binary separation, $a$, were both scaled to unity. The binary orbital period, $P_{orb}$, was scaled to $2\pi$. The radiation source was also scaled such that

\begin{equation}\label{eq:18}
\frac{L_{*Physical}}{L_{*Code}}=\frac{1}{2}G^{2/3} \left( \frac{2 \pi M_{t}}{P_{orb}}\right)^{5/3}.
\end{equation} 

\noindent Tables 1 \& 2 give details of the systems simulated.

\begin{table}
\label{table:DiscParameters}
\begin{center}
\caption
  {
    Binary system parameters for all the systems modelled. The columns are: 1 the total system mass in solar masses, 2 the system orbital period in days and 3 the rate of mass loss from the secondary in solar masses per year. 
  }
\begin{tabular}{|c|c|c|c|}
\hline
Parameter & $M_t$       & $P_{orb}$  & $\dot{M}_{sec}$    \\
Units     & $M_{\odot}$ & $days$     & $M_{\odot}yr^{-1}$ \\
\hline
Value     & 0.685       & $0.063121$ & $1.0\times10^{-10}$ \\

\hline
\end{tabular}
\end{center}
\end{table}

\begin{table*}
\label{table:ModelParameters}
\begin{center}
\caption
  {
    Parameters for the systems modelled. The columns are: the model run number, the system mass ratio, the physical luminosity and the ratio of the physical luminosity to the Eddington limit luminosity.  
  }
\begin{tabular}{|c|c|c|c|}
\hline
Model & q &  Physical luminosity ($L_{*}$) & $L_{*}/L_{Edd}$\\
number& ($M_{2}/M_{1}$) & ($erg\ s^{-1}$) & ($L_{*}/1.3\times10^{38}(M_{1}/M_{\odot})$) \\

\hline
 1 & 0.075 & $2.071\times10^{36}$ & 0.025 \\
 2 & 0.075 & $4.142\times10^{36}$ & 0.050 \\
 3 & 0.075 & $8.284\times10^{36}$ & 0.100 \\
 4 & 0.100 & $2.024\times10^{36}$ & 0.025 \\
 5 & 0.100 & $4.048\times10^{36}$ & 0.050 \\
 6 & 0.100 & $8.095\times10^{36}$ & 0.100 \\
 7 & 0.200 & $1.855\times10^{36}$ & 0.025 \\
 8 & 0.200 & $3.710\times10^{36}$ & 0.050 \\
 9 & 0.200 & $7.421\times10^{36}$ & 0.100 \\
10 & 0.300 & $1.713\times10^{36}$ & 0.025 \\
11 & 0.300 & $3.425\times10^{36}$ & 0.050 \\
12 & 0.300 & $6.850\times10^{36}$ & 0.100 \\
13 & 0.400 & $1.590\times10^{36}$ & 0.025 \\
14 & 0.400 & $3.180\times10^{36}$ & 0.050 \\
15 & 0.400 & $6.361\times10^{36}$ & 0.100 \\
16 & 0.075 & $4.142\times10^{37}$ & 0.500 \\
17 & 0.400 & $3.180\times10^{37}$ & 0.500 \\

\hline
\end{tabular}
\end{center}
\end{table*}

The accretion disc had an open inner boundary condition in the form of a hole of radius $r_{1}=0.025a$ centered on the position of the primary object. Particles entering the hole were removed from the simulation. Particles that re-entered the secondary Roche lobe were also removed from the simulation as were particles that were ejected from the disc and had a distance $> 0.9a$ from the centre of mass of the primary. 

We assumed a locally isothermal equation of state and that the dissipation was radiated from the point at which it was generated.  The Shakura-Sunyaev (1973) viscosity parameters were set to $\alpha_{low}$ of 0.01 and $\alpha_{high}$ of 0.1, and the viscosity state changed smoothly as described in Truss et al. (2000). The SPH smoothing length, $h$, was allowed to vary in both space and time and had a maximum value of $0.01a$, as described in Murray (1996, 1998). 

\subsubsection{The gas stream}
We simulated the mass loss from the secondary by introducing particles at the inner Lagrangian point ($L_1$). The mass transfer rate and the particle transfer rate were provided as input parameters,  and the mass of each particle was derived from these parameters. A particle was inserted with an initial velocity in the orbital plane according to the local sound speed of the donor in a direction prograde of the binary axis. The z velocity of the inserted particle was chosen from a Gaussian distribution, with a zero mean and a variance set to be a tenth of the local sound speed of the donor.

\subsubsection{The initial non-warped accretion disc}

The simulations were all started with zero mass in the accretion disc and with the radiation source switched off. A single particle was injected into the simulation every $0.01\Omega_{orb}^{-1}$ at the $L_1$ point as described above until a quasi-steady mass equilibrium was reached within the disc. This was taken to be when the number of particles inserted at the $L_1$ point, the mass transfer rate, was approximately equal to the number of particles leaving the simulation at the accretor, the accretion rate. The simulations were continued for another 3 orbital periods to ensure mass equilibrium. In all simulations the number of particles in the accretion disc was approximately 50,000 giving a good spatial resolution and the average number of `neighbours', that is the average number of particles used in the SPH update equations, was 8.9 particles. For the simulations with $q < 0.4$ the disc became eccentric and encountered the Lindblad 3:1 resonance. These discs precessed in a prograde direction giving rise to superhumps in the simulated dissipation light curves (c.f. Foulkes et al. 2004). 

The radiation source was then turned on and gave rise to a very small number of particles being ejected from the accretion disc. A better method would be to introduce the radiation slowly using a time exponential that gradually increased the intensity of the radiation over a period of time. However, the disc quickly recovered and continued as a normal disc for a short period until the radiation force generated a warp and tilted the disc out of the orbital plane.

\subsection{Surface finding algorithm \& self-shadowing}

To apply the radiation force equation (\ref{eq:17}), the particles on the surface of the accretion disc had to be identified. We implemented a convex hull algorithm to find the surface particles. The convex hull of a set of points is the minimum sub-set of points that completely enclose all other points. The algorithm was applied to the set of points $P = \left \{\mathbf{p}_1, \mathbf{p}_2, ... \right \}$ where $\mathbf{p}_i$ is the position vector of the particle $i$, and $i = 1,...,N$ where $N$ is the number of particles in the accretion disc simulation. The convex hull of the set $P$ is the smallest set $C \in P$ which encloses all of the points $\mathbf{p}_i$ (Clarkson et al. 1993).  The convex hull is analogous to a  rubber band stretched over the points. Care was taken to exclude particles that were far away from other particles and were clearly not part of the disc surface. The surface normal was then calculated using

\begin{equation}\label{eq:21}
\mathbf{n} = \left(\mathbf{p}_1-\mathbf{p}_2\right) \times \left(\mathbf{p}_2-\mathbf{p}_3\right)
\end{equation}

\noindent where $\mathbf{p}_1$,  $\mathbf{p}_2$ and $\mathbf{p}_3$ are adjacent surface element position vectors that form a surface triangle (i.e. do not lie on a straight line). 

The accretion disc surface was then constructed from the set of surface points. A ray-tracing algorithm was implemented which was used to determine regions of self-shadow. For each particle found on the disc surface a light-ray was projected from the particle to the position of the radiation source at the centre of the disc. The particle was deemed to be illuminated by the radiation source if this light-ray did not intersect any disc material between the particle surface position and the radiation source. The radiation force was only applied to particles that were considered to form part of the disc surface and were illuminated by the central radiation source.

\subsection{Disc warping and precession measure}

The two measures defined by Larwood \& Papaloizou (1997) were used to measure the disc warping and the amount of warp precession. They defined an angle $j$ as the angle between the total disc angular momentum vector and the angular momentum vector for a specific disc annulus, i.e.

\begin{equation}\label{eq:22}
\cos\ j = \frac{\mathbf{J}_A \cdot \mathbf{J}_D}{\big | \mathbf{J}_A \big | \big | \mathbf{J}_D \big |}.
\end{equation} 

\noindent The term $\mathbf{J}_A$ is the total angular momentum within the specific annulus and was calculated by summing the angular momentum for each particle within the annulus. The term $\mathbf{J}_D$ is the total disc angular momentum and was calculated by summing all the angular momenta for all particles within the disc. An angle $\Pi$ was also defined which measures the amount of precession of the disc angular momentum relative to the initial binary orbital angular momentum, $\mathbf{J}_O$

\begin{equation}\label{eq:23}
\cos\ \Pi = \frac{\left(\mathbf{J}_O \times \mathbf{J}_D\right) \cdot \mathbf{u}}
{\big | \mathbf{J}_O \times \mathbf{J}_D \big | \big | \mathbf{u} \big |}
\end{equation} 

\noindent where $\mathbf{u}$ is any arbitrary vector in the binary orbital plane. We also measured the tilt of the entire warped accretion disc relative to the orbital plane using

\begin{equation}\label{eq:24}
cos\ \beta = \frac{\mathbf{J}_O \cdot \mathbf{J}_D } {\big | \mathbf{J}_O \big | \big | \mathbf{J}_D \big | }.
\end{equation} 

\section{NUMERICAL RESULTS} 
\label{NUMERICAL RESULTS}

Before detailing our numerical results we briefly summarise our findings. In all the systems simulated, a warp developed and it precessed as a solid body in a retrograde direction relative to the inertial frame. After switching the radiation source on we ran each model for 50 orbital periods and the disc warp developed within five orbital periods after applying the radiation source. The rate of precession of the warp was related to the intensity of the radiation source, the higher the radiation intensity the more rapid the precession rate. The warped disc for the extreme mass ratio systems remained in the orbital plane whereas for less extreme mass ratio systems the whole disc also became inclined to the orbital plane. For the inclined discs the gas stream no longer impacted the edge of the disc and instead it interacted with the disc near the circularization radius for most of the binary orbit. Whenever we removed the radiation source the warp would dissipate and the disc would return to the orbital plane.

\subsection{Disc warping, models 16 and 17}

In this section we describe in detail the results from models 16 and 17, see Tables 1 \& 2. Model 16 had a mass ratio, ($q$), of $0.075$ and model 17 had a $q=0.4$. The non-irradiated accretion disc for model 16 was eccentric and precessed in a prograde direction relative to the inertial frame. The disc shape changed over an orbital period and generated superhumps in the simulated dissipation light curve. For model 17 the disc was constant in size and shape. The simulated light curve was constant with respect to time. Both models had a radiation source strength of approximately $50\%$ of the Eddington limit. We have placed Microsoft AVI movies of these warped discs on an Internet web-site (see http://physics.open.ac.uk/FHM\_warped\_disc/ for more information).

\subsubsection{Model 16 }

\begin{figure}
  \psfig{file=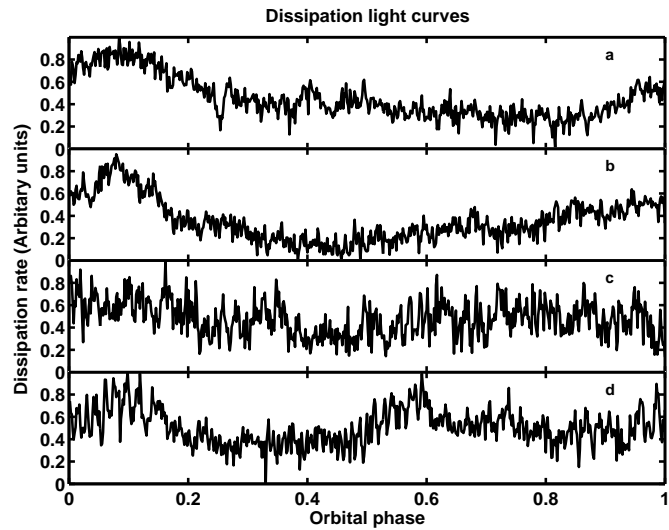,width=0.5\textwidth,angle=0}
  \caption{   
	         Simulated dissipation light curves for models 16 \& 17. The vertical axis in each plot has been normalised. The horizontal axis is time for a complete orbital period. The dissipation is for the whole disc and was computed by summing all particle dissipation for each time step. Plot (a) is the light curve for model 16 with the central radiation source switched off and plot (b) is with the radiation source switched on. Plots (c) and (d) are the light curves for model 17 without and with the central radiation source. 
         }
  \label{figure:lightcurves}
\end{figure}

Fig. \ref{figure:lightcurves} (a) shows the superhump in the simulated dissipation light curve for this model with the radiation source turned off. The light curve was generated by summing the dissipation for each particle over the whole of the disc for a complete orbital period. The modulation in this light curve is a positive superhump that repeated on a period a few percent longer than the orbital period. The superhump period is the period on which the donor star position repeats relative to the progradely precessing accretion disc. The gas stream impacts the outer rim of the disc and the energy dissipation light curve is dominated by the energy released by this process. 

When the radiation source was switched on a strong warp developed close to the centre of the disc. The simulated dissipation light curve for the warped disc is shown in Fig. \ref{figure:lightcurves} (b). The dissipation curves with and without the radiation source were very similar in amplitude and shape indicating that the outer edge dynamics of this disc were only marginally affected by the radiation source. 

Negative superhumps have a period that is a few percent shorter than the orbital period. Patterson et al. (1993) postulates that the precession of a warp within a disc can generate negative superhumps. In this simulation the precession of the warp relative to the binary frame does not add this type of cyclic component to the energy dissipation. Hence in this case at least a warped precessing disc is present but does not cause dissipation-powered negative superhumps.

\begin{figure*}
  \psfig{file=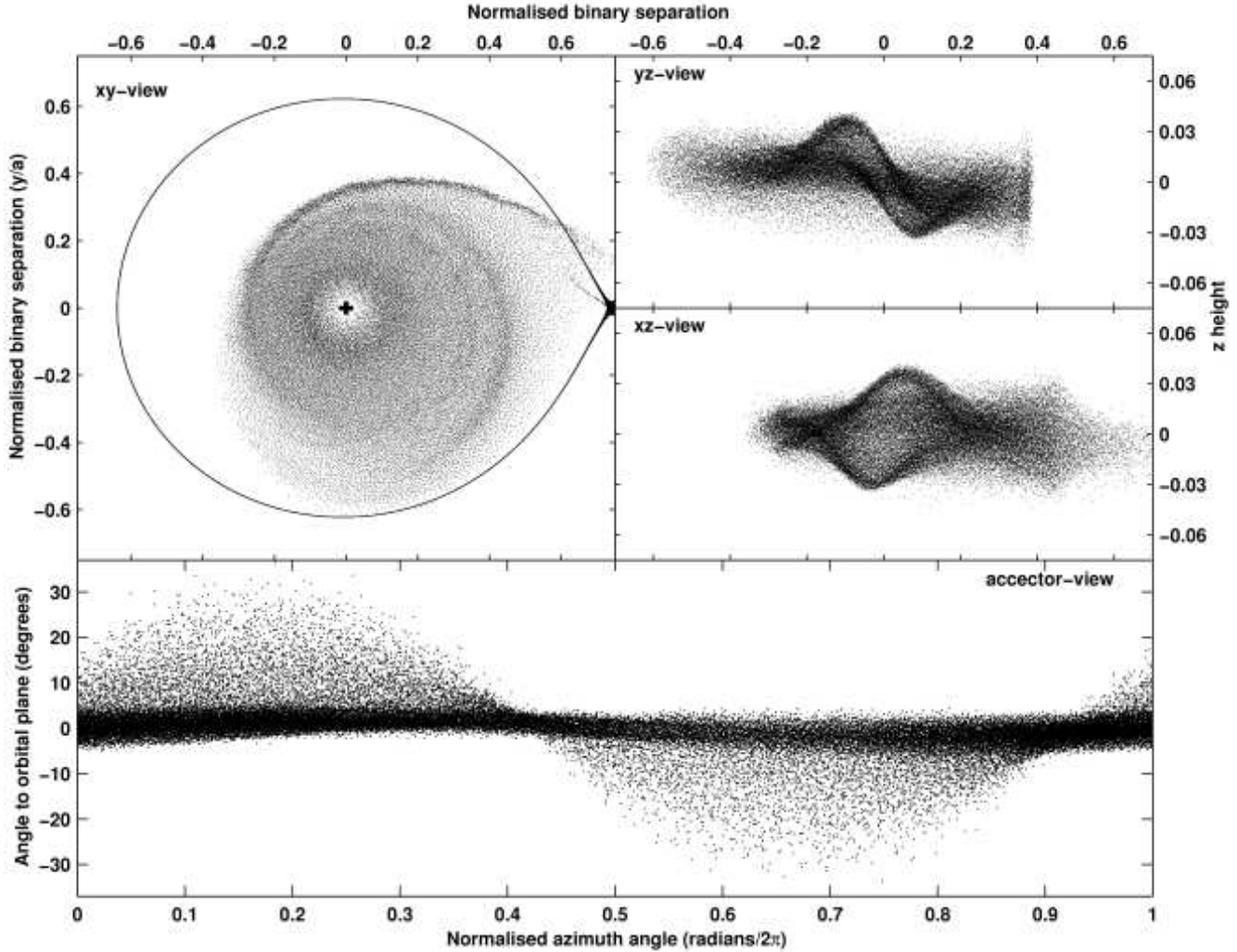,width=1.0\textwidth,angle=0}
  \caption{   
	   Particle projection plots for model 16. The position of each particle is indicated by a small black dot. The plot labelled  xy-view is a plan view of the accretion disc as seen from above the disc. The cross at the centre of the plot shows the position of the primary object. The solid dark line is the Roche lobe of the primary and the $L_1$ point is to the right and middle of the plot. The two plots xz-view and yz-view are particle projection plots on a plane perpendicular to the orbital plane and through the system axis. The bottom plot, accretor-view, shows the particle distribution as seen from the compact object. The horizontal axis is the normalised azimuth angle, the $L_{1}$ point is at angle 0 and the stream/disc impact region is at approximately angle 0.9. The vertical axis is the angle, in degrees, between a particle and the orbital plane when viewed from the compact object. The disc material flows from right to left.
         }
  \label{figure:figure_1}
\end{figure*}

Fig. \ref{figure:figure_1} contains projection plots for model 16; in all plots the position of each particle is indicated by a small black dot.  The upper left hand plot, labelled xy-view, is a plan view of the accretion disc as seen from above the disc. The solid dark line is the Roche lobe of the primary and the $L_1$ point is to the right. Material from the secondary enters the primary Roche potential from the $L_1$ point. The cross at the centre of the plot is the position of the primary object. A very strong spiral density compression wave can be seen at the upper edge of the disc. This wave is so intense that it is removing material from the accretion disc and returning it back to the Roche lobe of the secondary, see Foulkes et al. (2004) for a full detailed description of a 2D simulation of similar behaviour in a system with a mass ratio of $0.1$. 

The two upper right-hand plots of Fig. \ref{figure:figure_1}, labelled yz-view and xz-view, are side views of the disc in the y-z and x-z directions respectively. The yz-view plot is a projection view of the disc as seen from the secondary, similarly the xz-view is a projection plot with the secondary located to the right of the plot.  The disc warp is clearly apparent in these two plots. The warp is odd symmetrical about the centre of the disc. The maximum value of the warp is located at a distance approximately $0.1a$ either side of the primary position, see yz-view of Fig. \ref{figure:figure_1}.   

The lower plot of Fig. \ref{figure:figure_1}, labelled accretor-view, shows the distribution of the particles as seen from the compact object. The horizontal axis is the normalised azimuth angle, $L_1$ is located at angle $0$ and disc material flows from right to left with the stream-disc impact region located at approximately angle $0.9$. The vertical axis is the elevation angle of the particle as seen from the primary position. From this plot it can be seen that the radiation force has pushed disc material out of the orbital plane. The warp has a maximum value above the orbital plane at approximately an angle of $0.18$ and a minimum value at approximately angle $0.68$. This plot also shows that for this system the disc remains mainly in the orbital plane, although it can be seen that there is a small S-wave in the structure of the disc.

\begin{figure}
  \psfig{file=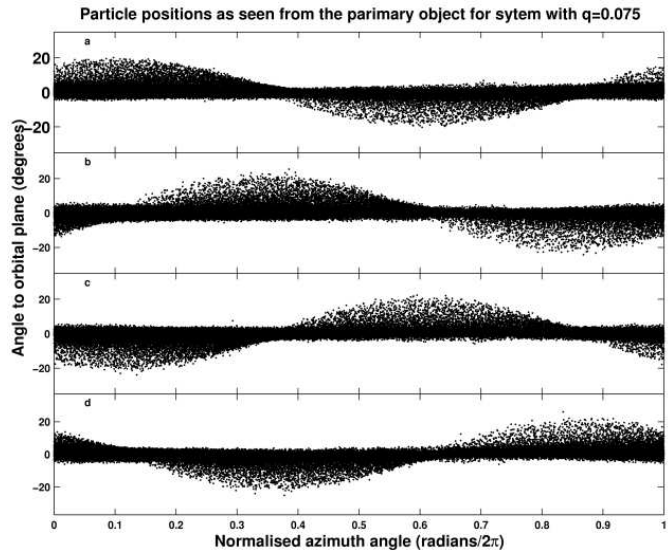,width=0.5\textwidth,angle=0}
  \caption{   
	   Model 16 accretor-view plots that show the particle distribution as seen from the accretor. The horizontal axis is the normalised azimuth angle, the $L_{1}$ point is at angle 0 and the stream/disc impact region is at approximately angle 0.9. The vertical axis is the angle, in degrees, between a particle and the orbital plane when viewed from the accretor. The disc material flows from right to left. The plots are four different instants separated by two orbital periods showing the warp propagating in a retrograde direction.
         }
  \label{figure:prec_model16}
\end{figure}

The warp precessed as a solid body in a retrograde direction relative to the inertial frame. The precession rate per orbital period was approximately $45^o$. The warp shape was very stable over many orbital periods; we ran this simulation for 50 orbital periods and the warp shape remained mostly constant throughout. Fig. \ref{figure:prec_model16} shows accretor particle plots for different orbital snapshots, sampling a complete warp precession. The warp can be seen to be traveling in a retrograde direction from left to right. The maximum amplitude of the warp in Fig. \ref{figure:prec_model16} (a) is located at approximately angle $0.1$; it travels to angles $0.35$, $0.6$ and $0.85$ in  Fig. \ref{figure:prec_model16} (b), (c) and (d) respectively. 

For an external observer viewing the central source at a high inclination, the radiation source will be obscured as the warp material comes between the observer and the radiation source.  The radiation will turn on and off on a period determined by the beat period between the orbital period and the retrograde precession rate of the disc warp. Consequently a negative superhump may be observed for systems at $\geq$ \degree{80} inclinations.

\subsubsection{Model 17}
\begin{figure*}
  \psfig{file=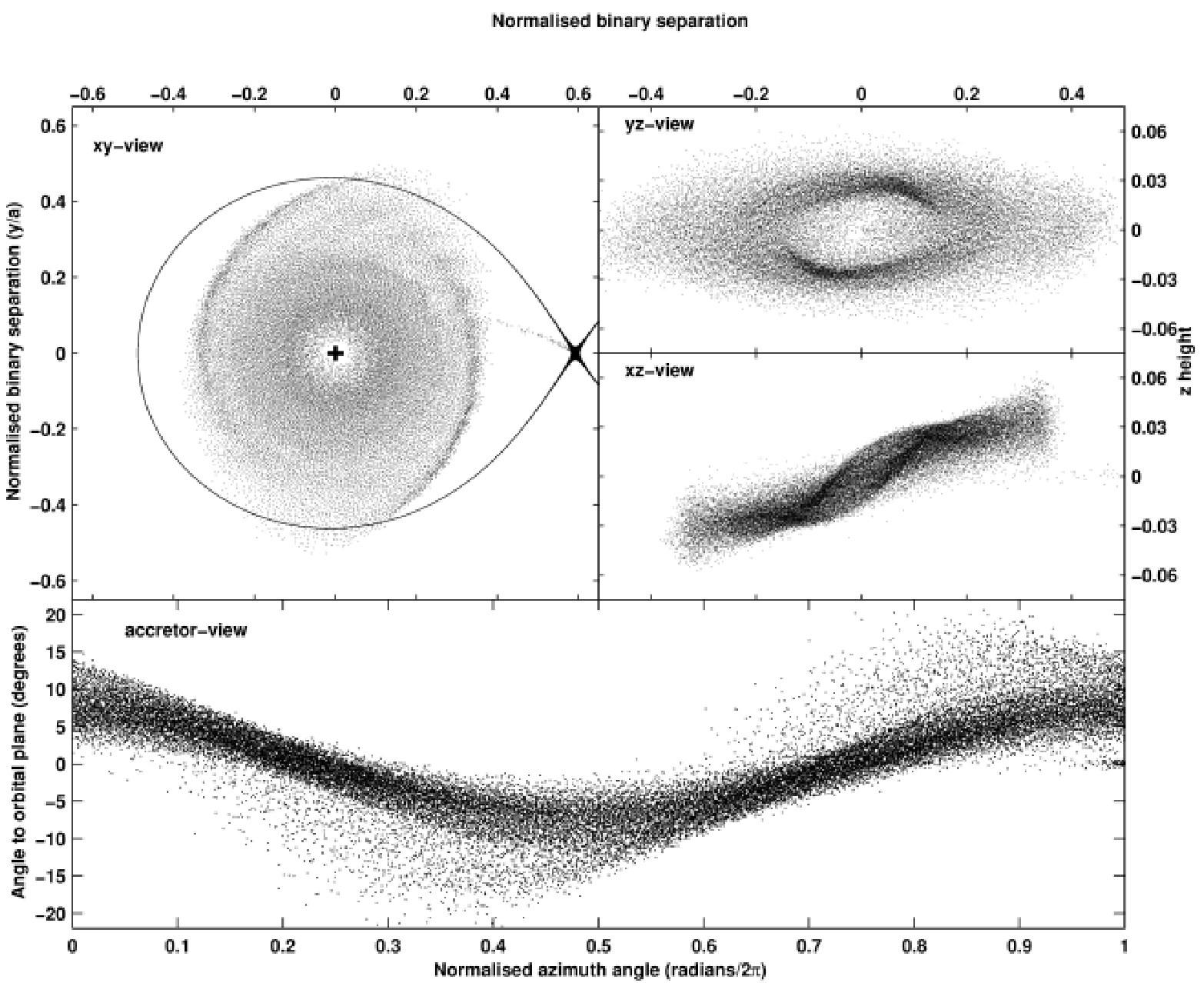,width=1.0\textwidth,angle=0}
  \caption{   
	   Particle projection plots for model 17. See Fig. \ref{figure:figure_1} for a description of the plots. The xz-view shows the disc almost edge on and the gas stream can be seen missing the edge of the disc.
         }
  \label{figure:figure_2}
\end{figure*}

The accretion disc for model 17 without the radiation source was roughly point-symmetric about the compact object. The disc did not precess and remained approximately constant in shape and size once mass equilibrium had been reached. Fig. \ref{figure:figure_2} shows projections (as described in Fig. \ref{figure:figure_1}) for model 17. The disc had two spiral density compression waves at the edge of the disc. The plan view of the disc remained approximately constant throughout the run.

\begin{figure}
  \psfig{file=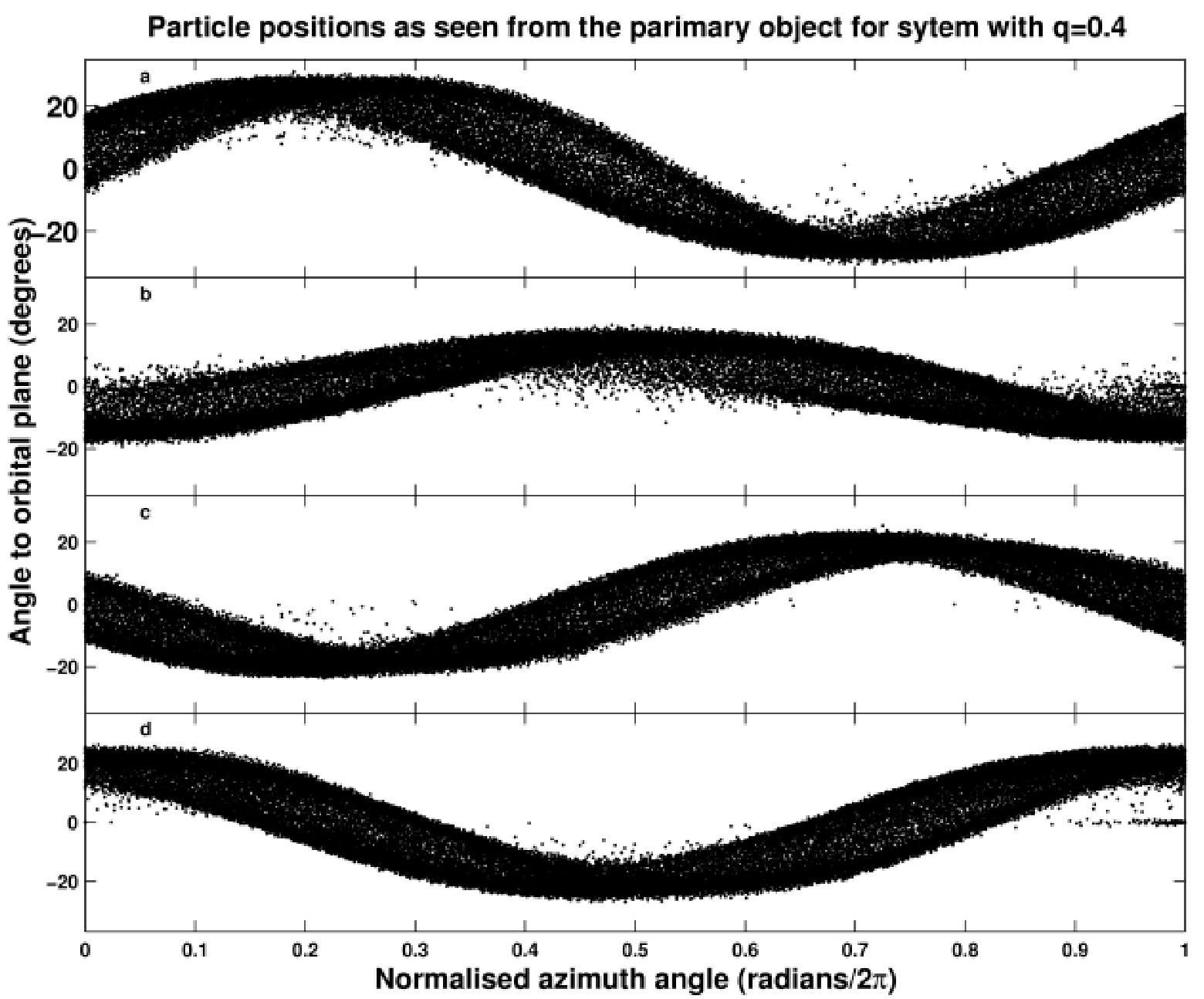,width=0.5\textwidth,angle=0}
  \caption{   
	   Model 17 accretor-view plots see caption of Fig. \ref{figure:prec_model16}. The plots are four different instants separated by six orbital periods showing the entire disc is tilted and propagating in a retrograde direction.
         }
  \label{figure:prec_model17}
\end{figure}

The two upper right-hand plots of Fig. \ref{figure:figure_2}, show that this disc was inclined with respect to the orbital plane. In the x-z view the disc is viewed almost side on. The disc warp can seen either side of the position of the accretor. The warp is odd symmetrical about the centre of the disc. The maximum value of the warp is located at a distance of approximately $0.15a$ either side of the primary position. The lower plot of Fig. \ref{figure:figure_2}, labelled accretor-view, shows the distribution of the particles as seen from the accretor. From this plot it can be seen that the radiation force has pushed disc material out of the orbital plane. This plot also shows that for this system the entire disc has tilted out of the orbital plane. The inclination of the disc remained constant for many orbital periods and the disc precessed in a retrograde direction, see Fig. \ref{figure:prec_model17}. The period of the retrograde precession was approximately $24$ orbital periods. We ran this simulation for 50 orbital periods and the tilt and warp shape remained approximately constant throughout this period. 

The simulated dissipation light curve for this system without the radiation source is shown in Fig. \ref{figure:lightcurves} (c). The curve shows little variation and is essentially constant except for a stochastic high frequency random noise. The dissipation light curve for the warped disc is shown in Fig. \ref{figure:lightcurves} (d). This curve is very different from the non-warped curve having two distinct humps at orbital phase $0.1$ and $0.6$. The period of these humps is a few percent shorter than the orbital period and is approximately equal to $(1-0.041)P_{orb}$. The period of the humps is the beat period of the orbital period and the retrograde precession period of the tilted disc. In this simulation, therefore, we see dissipation-powered negative superhumps.

The warped disc is inclined relative to the orbital plane as shown in Fig. \ref{figure:figure_2} which changes the point at which the gas stream impacts the accretion disc. For a non-warped accretion disc the material from the gas stream always impacts the outer edge of the disc at approximately the same point and energy dissipated at this point is constant with time. For the warped disc this is not the case. As the secondary orbits the tilted disc, the impact point of the gas stream varies with time. This impact point will change position from the edge of the disc to nearer its centre. As the gas stream material follows a ballistic path in the gravitational potential of the primary object it gains kinetic energy which is released at the disc impact point. The two light curve maxima correspond to when the disc faces toward or away from the secondary star. At this orbital phase the gas stream material will have maximum kinetic energy when it impacts on the disc. When the impact point is on the edge of the disc, the gas stream material will have a lower kinetic energy and this will correspond to the minima in the light curve.

The radiation source will be hidden to a high inclination observer as the tilted disc moves between the observer and the radiation source. This will occur at two points during the disc retrograde precession period. The radiation may also be obscured by the material in the warp close to the centre of the disc.

\subsubsection{Models 1-15}

In this subsection we summarise the results for models 1-15. All the simulations developed a warp that precessed in a retrograde direction. 

\begin{figure}
  \psfig{file=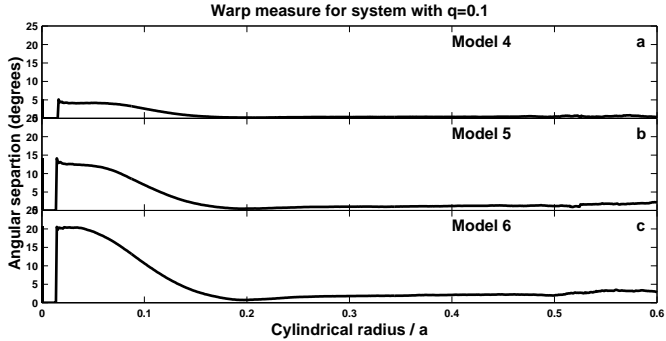,width=0.5\textwidth,angle=0}
  \caption{   
	   Warp profiles evaluated using equation (\ref{eq:22}). The plots are for a system with a mass ratio of 0.1. The vertical axis is the warp amplitude. The horizontal axis is distance from the primary object normalised such that the binary separation is 1. The three models have the same parameters except for the luminosity, which increased by a factor of 2 between successive models.  
         }
  \label{figure:warp_plot_one_system}
\end{figure}

We measured the warp induced by the radiation pressure using the angle defined in equation (\ref{eq:22}). Fig. \ref{figure:warp_plot_one_system} shows the warps generated for a single system, $q=0.1$ with different radiation intensities, (a) low intensity, (b) medium and (c) the highest intensity. In all three cases the warp is close to the primary object and increases with increasing radiation intensity. Although the warp amplitude increases with increasing luminosity, the warp region remains approximately constant, in all three plots the warp does not extend more than $\sim0.2a$. In this model the outer regions of the disc are precessing in a prograde direction which acts to damp the radial extent of inner warp. 

\begin{figure}
  \psfig{file=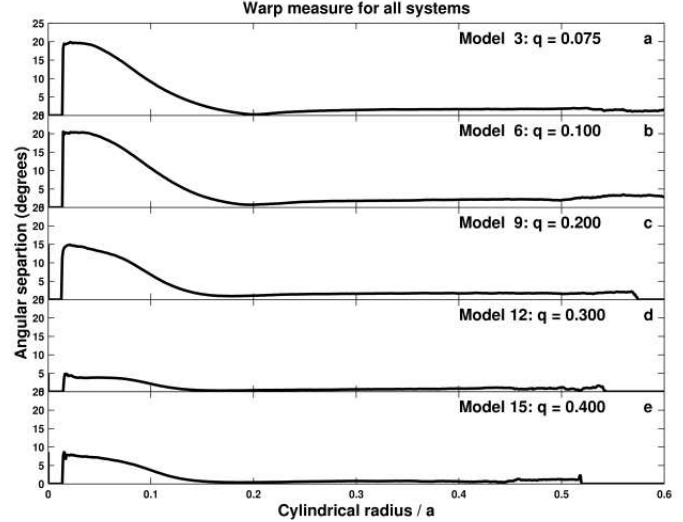,width=0.5\textwidth,angle=0}
  \caption{   
	   Warp profiles evaluated using equation (\ref{eq:22}). The plots are for a constant luminosity of approximately $10\%$ $L_{Edd}$. The vertical axis is the warp amplitude. The horizontal axis is distance from the primary object normalised so that the binary separation is 1. Plot (a), (b), (c), (d) and (e) correspond to models 3, 6, 9, 12 and 15 respectively.  
         }
  \label{figure:warp_all_systems}
\end{figure}

Fig. \ref{figure:warp_all_systems} shows the dependence on mass ratio of the warp profile. The luminosity was constant at approximately 10\% $L_{Edd}$. Fig. \ref{figure:warp_all_systems} (a), (b), (c), (d) and (e) correspond to mass ratios of $0.075$, $0.1$, $0.2$, $0.3$ and $0.4$ respectively. The horizontal axis is the distance from the primary object and the vertical scale remains the same in all plots. The figure clearly shows that the magnitude of the warp is a function of the system mass ratio and that the inner warp amplitude is larger for smaller values of $q$. 

\begin{figure}
  \psfig{file=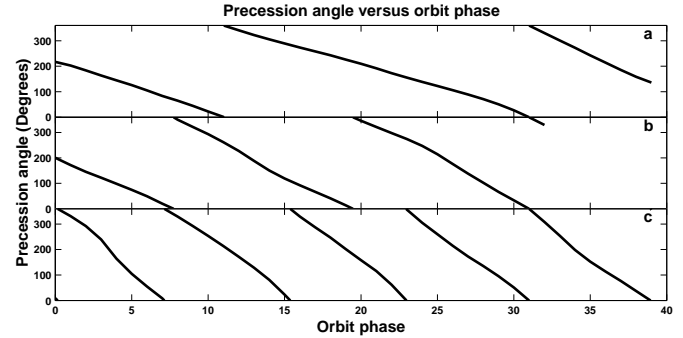,width=0.5\textwidth,angle=0}
  \caption{   
	   Precession angle as measured using equation (\ref{eq:23}) for a system with a mass ratio of $0.075$. The vertical axis shows the precession angle, in degrees, of the warp relative to some arbitrary start angle. The horizontal axis is the orbital phase. Plot (a) is for the system with central luminosity, $L_*$ equal to a value of $2.5\%$ $L_{Edd}$, plots (b) and (c) are for systems with central luminosities of $5\%$ and $10\%$ respectively.
         }
  \label{figure:prec_plot_one_system}
\end{figure}

\begin{figure}
  \psfig{file=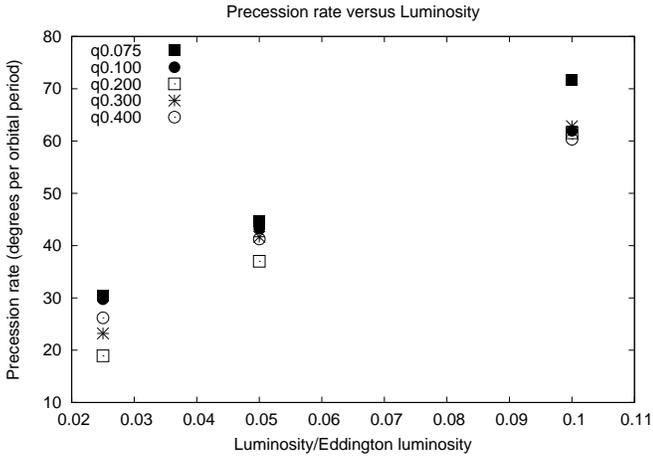,width=0.5\textwidth,angle=0}
  \caption{   
	   Accretion disc warp precession rate as a function of the central source luminosity for models 1-15, see key on the upper left hand side. The vertical axis is the precession rate of the disc measured in degrees per orbital period. The horizontal axis is the physical luminosity/Eddington limit.
         }
  \label{figure:prec_plot_all_systems}
\end{figure}

The rate of precession of the warp was determined by applying equation (\ref{eq:23}) to each time step of the simulations. Fig. \ref{figure:prec_plot_one_system} shows the precession angle versus orbital phase for models $1$, $2$, and $3$ in which the luminosity increased twofold between successive models. The three plots in this figure show how the precession rate varied with time. The warp precession rate was determined by fitting a straight line to the data using a Numerical Recipes least squares method, Press et al. (1986). The precession rate was found using the gradient of each line extracted from the least squares fits. In Fig. \ref{figure:prec_plot_all_systems} we show the precession rate in degrees per orbit versus luminosity for models 1-15. The figure shows that the precession rate increases with increasing luminosity. The precession rate is  also related to the amplitude for the inner warp. 

\begin{figure}
  \psfig{file=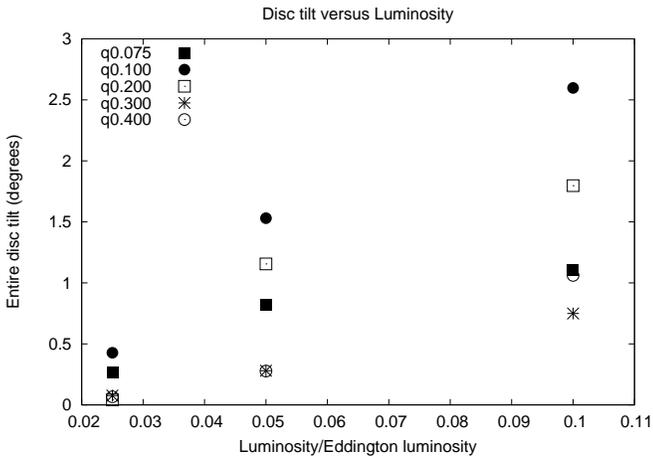,width=0.5\textwidth,angle=0}
  \caption{   
	   Accretion disc tilt angle as a function of the central source luminosity for all the systems simulated, see key on the upper left hand side. The vertical axis is the final tilt angel of the disc calculated using equation (\ref{eq:24}), measured in degrees. The horizontal axis is the physical luminosity/Eddington limit. 
         }
  \label{figure:disc_tilt}
\end{figure}

The tilt of the accretion disc, $\beta$, was calculated using equation (\ref{eq:24}) for each system simulated and the results are displayed in Fig. \ref{figure:disc_tilt}. The tilt angle of the disc increased with increasing radiation pressure. The figure also indicates that for extreme mass ratios the disc has a small inclination and remains mostly in the orbital plane. For these systems a large warp close to the primary object develops and precesses in a retrograde direction. For less extreme mass ratio systems a smaller warp develops close to the primary and the disc has a larger tilt angle out of the orbital plane which then precess as a solid body in a retrograde direction.

\begin{figure}
  \psfig{file=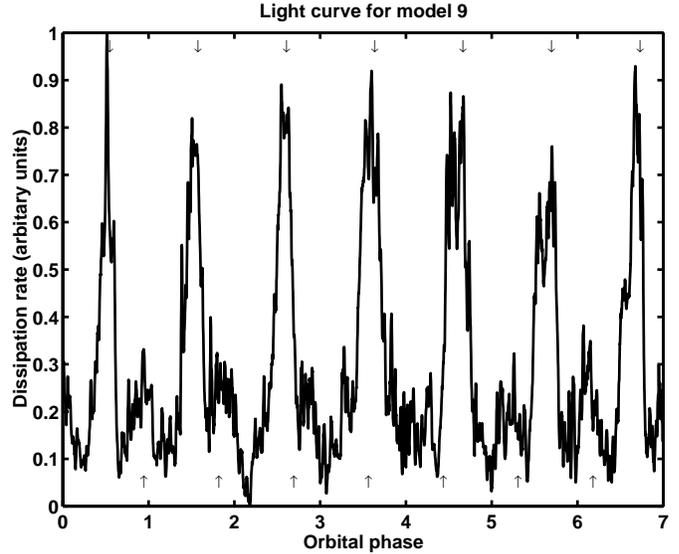,width=0.5\textwidth,angle=0}
  \caption{   
	    Simulated dissipation light curve for model 9, the vertical axis in the plot has been normalised. The horizontal axis is time measured in orbital periods. The dissipation is for the whole disc and was computed by summing all particle dissipation for each time step. The downward pointing arrows at the top of the plot indicate the position of the positive superhumps. The upward pointing arrows at the bottom of the plot indicate the position of the negative superhumps. 
         }
  \label{figure:light_curve_q_0p2}
\end{figure}

In Fig. \ref{figure:light_curve_q_0p2} we show the disc dissipation light curve for a system with $q=0.2$ and a high central luminosity, model 9 $L=0.1L_{Edd}$. The disc is just tilted out of the orbital plane and the entire disc precesses slowly in a retrograde direction with respect to the orbit. The large peaks in this plot are positive superhumps and have a period slightly longer than the orbital period and are indicated by the downward pointing arrows at the top of the plot. The positive superhumps are a result of the interaction of the gas stream and the edge of the disc which rotates in a prograde direction. The period of the superhumps is the period which the secondary crosses the line of apsides of the prograde precessing eccentric disc.  

The dissipation light curve in Fig. \ref{figure:light_curve_q_0p2} also shows a smaller modulation with a period slightly shorter than the orbital period and is indicated by the upward pointing arrows near the bottom of the plot. Since the tilted disc is precessing in a retrograde direction the secondary reaches conjunction with the line of nodes marginally  before it completes a full orbital period.  The period of the smaller modulation is then the beat period of the orbital period and the retrograde period of the tilted disc. As the secondary orbits the compact object the point at which the gas stream impacts the tilted disc varies. When the disc is tilted towards or away from the $L_{1}$ point some of the gas stream flows over or under the edge of the disc. The impact point of this material will change from the edge of the disc to nearer the compact object. The overflow material gains kinetic energy as it travels towards the central object which is subsequently released as dissipation when it hits the disc surface.  Hence for a system in which the disc is tilted out of the orbital plane while the outer disc precesses in a prograde direction, both positive and negative superhumps maybe observed simultaneously. Patterson et al. (1997) reported such simultaneous positive and negative superhumps in the CV V603 Aql.

\section{Conclusions}
\label{sec:Conclusions}  

\subsection{Algorithms}
We have studied the radiation-driven warping of accretion discs in X-ray binaries using a non-linear smoothed particle hydrodynamics code. A convex hull algorithm was developed to find the surface particles of the disc and a ray-tracing algorithm used to evaluate regions of self-shadowing. Initially planar accretion discs were illuminated with radiation from the centre of the discs. The surface of each disc was subject to a force from the absorption of the X-rays on the disc surface and back reaction from the remitted radiation. 

\subsection{Warped disc structure}
Our results indicate that twists and warps develop which undergo near rigid body precession in a retrograde direction relative to the inertial reference frame. These warps survive for many orbital periods in close binary systems, with the integrity of their local vertical structure maintained. The entire disc may also become inclined relative to the orbital binary plane.

We found that for the extreme mass ratio systems, a strong warp develops close to the centre of the disc to the extent that some of the outer parts are completely shadowed by the inner warp. This warp shadow and the viscous forces combine to keep the disc in the orbital plane and the inclination of the disc to the orbital plane is small. The warp precesses in a wave-like manner in a retrograde direction relative to the inertial frame. The disc as a whole still precesses in a prograde direction similar to a non-warped disc and superhumps are still apparent in the simulated dissipation light curve (c.f. Foulkes et al.).

For the less extreme mass ratio systems a small warp is induced in the inner regions of the disc. This warp does not shadow the outer regions of the disc and these regions also warp. The entire disc then becomes tilted to the binary plane and the whole disc precesses in a retrograde direction. The warp behaviour we find appears therefore to be consistent with the differing observational characteristics of the extreme mass ratio system X\thinspace 1916-053 and the non-extreme q system Her \thinspace X-1.

\subsection{Positive and negative superhumps}
In the less extreme mass ratio systems, as the tilt of the disc becomes larger the gas stream no longer impacts on the edge of the disc. Instead the stream now impacts deeper in the disc near the circularization radius and the simulated dissipation light curve shows two peaks per orbital period occuring when the stream-disc impact point is deepest in the gravitational well. 

Systems that have eccentric accretion discs in which the edge of the disc precesses in a prograde direction and the entire disc is inclined to the orbital plane, positive and negative superhumps are visible in the simulated disc dissipation light curves. The positive superhumps are generated by the interaction of the edge of the disc and the material from the gas stream. Since the edge material in the disc precesses in a prograde direction the period of the positive superhumps is slightly longer than the orbital period. When the whole of the disc is also marginally inclined out of the orbital plane and the entire disc warp is precessing in a retrograde direction, then negative superhumps are generated by some of the gas stream material over or under flowing the edge of the disc. The period of the negative superhumps is slightly shorter than the orbital period.

\subsection{Radiation field strength and behaviour at $L \sim L_{Edd}$}

The radiation field strength throughout the disc was measured using the dimensionless parameter $F*$ as defined by Wijers \& Pringle (1999) (their equation (8)) 

\begin{equation}\label{eq:f*}
F* = \frac{L_{*}}{6 \pi c r \Sigma \Omega \nu}.
\end{equation} 

\noindent where $L_{*}$ is central source luminosity, $c$ is the speed of light, $r$ is the distance from the compact object,  $\Sigma$ is the disc surface density at radius $r$, $\Omega$ is the angular velocity at radius $r$ and $\nu$ is the shear viscosity. This dimensionless number gives the viscous time scale at a point in the disc divided by the radiation time scale. Wijers \& Pringle (1999) found that $F*$ had to be greater than a critical value for warping of the disc. 

\begin{figure}
  \psfig{file=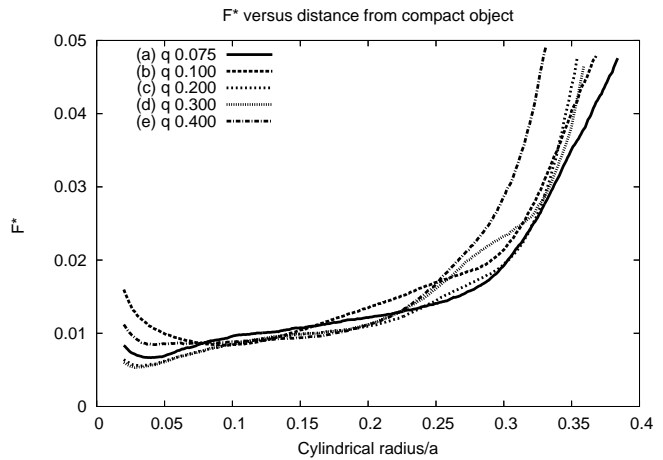,width=0.5\textwidth,angle=0}
  \caption{   
	    Radiation field strength evaluated using equation (\ref{eq:f*}). The plots are for a constant luminosity of approximately $2.5\%$ $L_{Edd}$. The vertical axis is the dimensionless radiation field strenght and the horizontal axis is distance from the primary object normalised so that the binary separation is 1. Plot (a), (b), (c), (d) and (e) are correspond to models 1, 4, 7, 10 and 13 respectively. The extent of each plot was calculated using a ray-tracing algorithm and shows the maximum disc radius that was illuminated by the central radiation source.
         }
  \label{figure:fstar}
\end{figure}

Fig. \ref{figure:fstar} shows $F*$ as a function of distance from the compact object for each system considered and for a luminosity equal to $2.5\%L_{Edd}$.  We used the ray-tracing algorithm to calculate the extent of the disc illuminated by the radiation source and the figure plots $F*$ out to this distance for each disc. The figure indicates that $F*$ is largest at the edge of the disc and that the disc warp starts at the edge of the disc and propagates towards the central object.

We conducted a number of simulations in order to investigate the warp behaviour as the central luminosity approached $L_{Edd}$. In our simulations whenever the central luminosity was just slightly larger than the Eddington limit the accretion disc material would start moving away from the central object and accretion at the central object would cease. The warp angles in our simulations never exceeded \degree{90} and were about \degree{30} - \degree{40} at most, even when the central luminosity was close to the Eddington limit. 

\subsection{Warp wave transportation process}
Fig. \ref{figure:fstar} indicates that the outer regions of the accretion disc that are illuminated by the central radiation source are most unstable to warping. The warp starts in these outer regions of the disc and propagates toward the central regions of the disc in the form of a prograde spiral. Wijers and Pringle (1999) indicate two processes for the transportation of a warp through a disc. The first process is a wave-like effect \cite{LubowPringle:1993,KorycanskyPringle:1995,PapaloizonLin:1995,OgilvieLubow:1999} and this regime was used by Larwood $\&$ Papaloizon (1997) for modelling warped circumbinary discs using a SPH code. The second process involves the exchange of angular momentum in a radial direction via viscous processes (Pringle 1992). Wijers and Pringle (1999) also indicate the condition for warp waves to propagate over a radial distance $R$ (their equation 30)

\begin{equation}\label{eq:wave}
\alpha \ll  \frac{H}{R},
\end{equation} 

\noindent where $\alpha$ is the normal Shakura-Sunyaev (1973) viscosity parameter and $H$ is the local disc thickness. In our current stimulations this condition was not satisfied and the warp transport process was dominated by viscous torques.

\subsection{Effects of donor star magnetic field}
Murray et al. (2002) modelled an initially planar disc subject to an inclined magnetic dipole field centred on the donor star. We conducted a limited number of simulations to investigate the warping effect reported and found that the warping was transient in nature. The warp developed strongly when the magnetic dipole field was first applied to the accretion disc but then dissipated after a period of time. Our number of particles was about a tenth of that used by Murray et al. (2002). 

\subsection{Future work}
In a future paper, already in preparation, we shall apply our SPH code to a number of systems that have photometric periodicities that are significantly longer than their orbital periods. In particular we will investigate Her X1/HZ X-1 and SS 433. Both of these systems have long periods in their observed optical and X-ray light curves which have been attributed to an inclined accretion disc that is precessing in a retrograde direction.

\section{Acknowledgments} SBF acknowledges the support from QinetiQ, Malvern. We acknowledge support from the Open University's Research School. Most of the runs were executed on The Swinburne Centre for Astrophysics and Supercomputing facility. We thank Jonathan Underwood \& Will Clarkson for comments on this work.

\label{lastpage}

\end{document}